\documentstyle[twoside,fleqn,epsfig]{article}

\newcounter{zacountsec} 
\newcommand{\eh}{\hfill}\newlength{\sperr} 
\newcommand{\Title}[1]{{\large \bf #1}} 
\newcommand{\Author}[1]{{\sc #1}} 
\newcounter{symposium}
\newcommand{\Symposium}[1]{\setcounter{symposium}{#1}} 
\newcounter{session}
\newcommand{\Session}[1]{\setcounter{session}{#1}} 
\expandafter\ifx\csname jtitle \endcsname\relax\def\jtitle{}\fi 
\expandafter\ifx\csname symptitle \endcsname\relax\def\symptitle{}\fi 
\expandafter\ifx\csname makeheadings \endcsname\relax\def\makeheadings{}\fi 
\newcommand{\Section}[1]{{\stepcounter{zacountsec}\vspace{3mm}%
\hspace*{18mm}\normalsize\bf\arabic{zacountsec}. \parbox[t]{150mm}{ #1 }}} 
\newenvironment{Abstract}{\begin{minipage}[t]{177mm}\em }
{\end{minipage}} 
\newenvironment{thm}[2]{\begin{sloppypar}
{#1 #2.}\em{}}
{\end{sloppypar}}
\newcommand{\proof}{\hspace*{9mm}{\settowidth{\sperr}{\rm Proof}%
\parbox[t]{1.3\sperr}{\rm P\eh r\eh o\eh o\eh f\eh. } }}
\newcounter{zalit}
\newenvironment{Acknowledgements}{\vspace{3mm}
\hspace*{18mm}{\bf Acknowledgements}\\[0.3cm]\begin{minipage}[t]{177mm}%
\small \em}{\end{minipage}}
\newenvironment{References}{
\Section{References}
\begin{small}\begin{list}{\arabic{zalit} }{\usecounter{zalit} 
\itemsep0mm \parsep0mm\settowidth{\labelwidth}{\small\rm 88}\labelsep0mm 
\setlength{\leftmargin}{\labelwidth}}}
{\end{list}\end{small}}
\newlength{\addrt}
\newenvironment{Address}{
\begin{minipage}[t]{\addrt}}
{\end{minipage}}
\newcommand*{\AuthorID}[1]{}\newcommand*{\LastName}[1]{}
\newcommand*{\FirstName}[1]{}\newcommand*{\ShortFirstName}[1]{}
\newcommand*{\Degree}[1]{}\newcommand*{\EMail}[1]{}
\newcommand*{\AuthorAddress}[1]{}
\newcommand{\AuthorInfo}[7]{%
\AuthorID{#1}\LastName{#2}\FirstName{#3}\ShortFirstName{#4}%
\Degree{#5}\EMail{#6}\AuthorAddress{#7}}
\newcommand{\ud}{\mathrm{d}}
\begin{document}
\AuthorInfo{1}
    {Fedorova}
    {Antonina}
    {A.}
    {}
    {anton@math.ipme.ru}
    {Russian Academy of Sciences, Institute of Problems of Mechanical Engineering,                    
     Mathematical Methods in Mechanics Group, V.O., Bolshoj pr., 61, 199178, 
     St.~Petersburg, Russia}
\AuthorInfo{2}
   {Zeitlin}
   {Michael}
   {M.}
   {Dr.}
   {zeitlin@math.ipme.ru}
   {Russian Academy of Sciences, Institute of Problems of Mechanical Engineering,                    
     Mathematical Methods in Mechanics Group, V.O., Bolshoj pr., 61, 199178, 
     St.~Petersburg, Russia}
\Symposium{0}

\Session{13} 
\makeheadings \markboth{\jtitle}{\symptitle}

\begin{minipage}[t]{180mm}
\thispagestyle{empty}
\vspace{20mm}

\begin{center}
{\Large\bf Localization and Coherent Structures}

\vspace{7mm}

{\Large\bf  in Wave Dynamics via Multiresolution} 

\vspace{20mm}

{\large\bf Antonina N. Fedorova, Michael G. Zeitlin}

\vspace{20mm}

Mathematical Methods in Mechanics Group \\

Institute of Problems of Mechanical Engineering (IPME RAS)\\ 

Russian Academy of Sciences \\

Russia, 199178, St. Petersburg, V.O., Bolshoj pr., 61\\

zeitlin@math.ipme.ru, anton@math.ipme.ru\\

http://www.ipme.ru/zeitlin.html\\

http://www.ipme.nw.ru/zeitlin.html

\vspace{20mm}
{\bf Abstract}

\vspace{10mm}

\begin{tabular}{p{100mm}}
We apply variational-wavelet approach for constructing multiscale high-localized eigenmodes expansions
in different models of nonlinear waves. We demonstrate appearance of coherent localized structures
and stable pattern formation in different collective dynamics models. 

\vspace{10mm}

Presented: GAMM Meeting, February, 2001, ETH, Z\"urich

\vspace{5mm}

Published: PAMM, Volume 1, Issue 1, pp. 399-400, Wiley-VCH, 2002

\end{tabular}

\end{center}
\end{minipage}
\newpage

\vspace*{15mm}\hspace*{18mm}
\begin{minipage}[t]{157mm}

\Author{Fedorova, A.; Zeitlin M.}
\vspace*{0.4cm}

\Title{Localization and Coherent Structures in Wave Dynamics via Multiresolution}
\end{minipage}

\vspace*{3.5mm}

\begin{Abstract}
We apply variational-wavelet approach for constructing multiscale high-localized eigenmodes expansions
in different models of nonlinear waves. We demonstrate appearance of coherent localized structures
and stable pattern formation in different collective dynamics models. 
\end{Abstract}

We consider the applications of a new nu\-me\-ri\-cal\--analytical 
technique based on the methods of local nonlinear harmonic
analysis or wavelet analysis to nonlinear wave dynamics problems.
Such approach may be useful in all models in which  it is 
possible and reasonable to reduce all complicated problems related with 
statistical/stochastic distributions to the problems described 
by systems of nonlinear ordinary/partial differential 
equations with or without some (functional) constraints (e.g. Kuramoto-Sivashinsky (KS) equation as a model
of weak turbulence).
Wavelet analysis gives us the possibility to work with well-localized bases in
functional spaces and gives for the general type of operators (differential,
integral, pseudodifferential) in such bases the maximum sparse forms. 
For KS equation
\begin{eqnarray}                                           
\Psi_t=-\Psi_{xxx}-\Psi_{xx}-\Psi \Psi_x
\end{eqnarray}          
and in related models we use the following variational approach.
Let L be an arbitrary (non) li\-ne\-ar (polynomial/rational) dif\-fe\-ren\-ti\-al\-/\-in\-teg\-ral operator with matrix dimension $d$, 
which acts on some set of functions
$\quad\Psi\equiv\Psi(t,x)=\Big(\Psi^1(t,x),\dots, \Psi^d(t,x)\Big)$, $ t,x\in\Omega\subset{\bf R}^{n+1}$
from $L^2(\Omega)$:
\begin{equation}
L\Psi\equiv L(Q,t,x)\Psi(t,x)=0, \qquad 
Q\equiv Q_{d_1,d_2}(t,x,\Psi,\partial /\partial t,\partial /\partial x)=\sum_{i_1,i_2}^{d_1,d_2}
a_{i_1i_2}(t,x.\Psi)
\Big(\frac{\partial}{\partial t}\Big)^{i_1}\Big(\frac{\partial}{\partial x}\Big)^{i_2}\nonumber
\end{equation}
Let us consider now the N mode approximation for solution as the following ansatz (in the same way
we may consider different ansatzes):
\begin{equation}
\Psi^N(t,x)=\sum^N_{r,k}a_{rk}A_r\otimes B_k(t,x), \qquad k\in Z^d
\end{equation}
We shall determine the coefficients of expansion from the following conditions
(different related variational approaches are considered in [1]-[4]):
\begin{equation}
\ell^N_{k\ell}\equiv\int(L\Psi^N)A_k(t)B_\ell(x) \ud t\ud x =0
\end{equation}
So, we have exactly $dN^{n+1}$ algebraical equations for  $dN^{n+1}$ unknowns $a_{rk}$.
Such variational approach reduces the initial problem to the problem of solution 
of functional equations at the first stage and some algebraical problems at the second
stage.
The solution is parametrized by solutions of two set of reduced algebraical
problems, one is linear or nonlinear
(depends on the structure of operator L) and others are some linear
problems related to computation of coefficients of algebraic equations (4).
These coefficients can be found  by some wavelet methods
by using
compactly supported wavelet basis functions for expansions (3).
The constructed solution has the following mul\-ti\-sca\-le\-/\-mul\-ti\-re\-so\-lu\-ti\-on decomposition via 
nonlinear high\--\-lo\-ca\-li\-zed ``eigenmodes'' 
{\setlength\arraycolsep{0pt}
\begin{eqnarray}\label{eq:z}
&&\Psi(t,x)=\sum_{(i,j)\in Z^{n+1}}a_{ij}A^i(t)B^j(x),\\
&&A^i(t)=A_N^{i,slow}(t)+\sum_{r\geq N}A^i_r(\omega^1_rs),\ \omega^1_r\sim 2^r ,\qquad
B^j(x)=B_M^{j,slow}(x)+\sum_{l\geq M}B^j_l(k_lx),\ k_l\sim 2^l \nonumber
\end{eqnarray}}
which corresponds to the full multiresolution expansion in all underlying time/space 
scales.
Formula (\ref{eq:z}) gives us expansion into the slow part $\Psi_{N,M}^{slow}$
and fast oscillating parts for arbitrary N, M. So, we may move
from coarse scales of resolution to the 
finest one to obtain more detailed information about our dynamical process.
The first terms in the RHS of formulae (5) correspond on the global level
of function space decomposition to  resolution space and the second ones
to detail space.
This representation (5) provides the solution as in linear as in nonlinear cases without any perturbation technique
but on the level of expansions in (functional) space of solutions.
The using of wavelet basis with high-localized properties provides good convergence properties of constructed solution (5). 
As a result of good (phase)space/time      
localization properties we can construct high-localized coherent structures in      
spa\-ti\-al\-ly\--ex\-te\-nd\-ed stochastic systems with collective behaviour.
In all these models numerical modelling demonstrates the appearance of coherent 
high-localized structures (Fig.1)
and stable patterns formation (Fig.2).

\begin{figure}[thb]
\centering
\epsfig{file=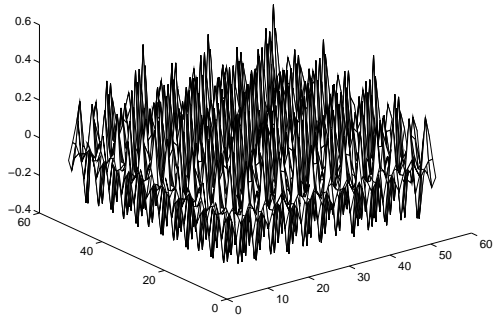, width=65mm}
\caption{Appearance of coherent structure}
\end{figure}

\begin{figure}[thb]
\centering
\epsfig{file=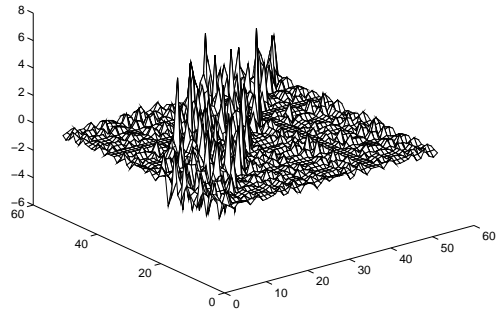, width=65mm}
\caption{Stable pattern}
\end{figure}

\begin{Acknowledgements}
We would like to thank ETH, Zurich for hospitality and support, 
which gave us the possibility to present our two papers during
GAMM 2001 Annual Meeting in Zurich and Prof. Martin Gutknecht
for permanent help and encouragement.
\end{Acknowledgements}

\begin{References}

\item {\normalsize \sc Fedorova, A., Zeitlin M.}:
Wavelets in Optimization and Approximations; 
Math. and Comp. in Simulation, {\bf 46} (1998), 527--534.

\item {\normalsize \sc Fedorova, A., Zeitlin M.}:
Variational Approach in Wavelet Framework to Polynomial 
Approximations of Nonlinear Accelerator Problems;
American Institute of Physics, CP, {\bf 468}, 
Nonlinear and Collective Phenomena in Beam Physics 
 (1999), 48--68.

\item  {\normalsize \sc Fedorova, A., Zeitlin M.}:
Variational-Wavelet Approach to RMS Envelope Equations;
The Physics 
of High Brightness Beams, World Scientific (2000), 235--254.

\item {\normalsize \sc Fedorova, A., Zeitlin M.}: 
Localized Coherent Structures and Patterns Formation 
in Collective Models of Beam Motion;
Quantum Aspects of Beam Physics, World 
Scientific (2001); Los Alamos preprint, physics/0101007.

\end{References}

\begin{Address}
{\sc Antonina Fedorova,}
Russian Academy of Sciences, 
Institute of Problems of Mechanical Engineering, \\                   
V.O., Bolshoj pr., 61, 199178, St.~Petersburg, Russia\\
email: anton@math.ipme.ru, http://www.ipme.ru/zeitlin.html,
http://www.ipme.nw.ru/zeitlin.html

\end{Address}

\end{document}